\documentclass[aps,prb,twocolumn,groupedaddress,showkeys,showpacs]{revtex4}

\usepackage{multirow,amssymb,amsbsy,amsmath}
\usepackage{epsfig}

\newcommand{\op}[1]{%
    \fontdimen12\textfont3=2pt\fontdimen12\scriptfont3=1.4pt%
    \!\null\mathop{\vphantom{#1}\smash{#1}}\limits_{\sim}\null\!}
\newcommand{\xref}[1]{\protect\ref{#1}}
\newcommand{\figref}[1]{Fig.~\protect\ref{#1}}
\newcommand{\fmref}[1]{(\protect\ref{#1})}

\begin{document}

\title{Magnetic characterization of the frustrated three-leg
  ladder compound [(CuCl$_2$tachH)$_3$Cl]Cl$_2$}

\author{J\"urgen Schnack}
\email{jschnack@uos.de}
\affiliation{Universit\"at Osnabr\"uck, Fachbereich Physik,
D-49069 Osnabr\"uck, Germany}

\author{Hiroyuki Nojiri}
\email{nojiri@imr.tohoku.ac.jp}
\affiliation{Institute for Materials Research, Tohoku
  University, Katahira 2-1-1, Sendai 980-8577, Japan}

\author{Paul K\"ogerler}
\email{kogerler@ameslab.gov}
\affiliation{Ames Laboratory \& Department of Physics and Astronomy,
Iowa State University, Ames, Iowa 50011, USA}

\author{Geoffrey J. T. Cooper}
\author{Leroy Cronin}
\email{L.Cronin@chem.gla.ac.uk}
\affiliation{Dept. of Chemistry, The University of Glasgow, Glasgow, G12 8QQ, UK}

\date{\today}

\begin{abstract}
  We report the magnetic features of a new one-dimensional stack
  of antiferromagnetically coupled equilateral copper(II)
  triangles. High-field magnetization measurements show that the
  interaction between the copper triangles is of the same order
  of magnitude as the intra-triangle exchange although only
  coupled via hydrogen bonds. The infinite chain turns out to be
  an interesting example of a frustrated cylindrical three-leg
  ladder with competing intra- and inter-triangle interactions.
  We demonstrate that the ground state is a spin singlet which
  is gaped from the triplet excitation.
\end{abstract}

\maketitle

%%%%%%%%%%%%%%%%%%%%%%%%%%%%%%%%%%%%%%%%%%%%%%%%%%%%%%%%%%%%%%%%%%%%%%%%
\section{Introduction}
\label{sec-1}

One characteristics of magnetic molecules is that their
intermolecular interaction is usually extremely weak compared to
their intramolecular interactions. This becomes one of the
experimental advantages when investigating these materials since
a measurement at a crystal or powder of such molecules reflects
the properties of a single molecule. Also for speculative
applications like storage media the independence of the
molecular units is crucial, for an overview see
Ref.~\onlinecite{GaS:ACIE03} and references therein.

Despite the advantages of virtually independent magnetic
molecules the controlled coupling of magnetic clusters to one-
or multidimensional networks is a highly interesting avenue as
well, especially if the degree of magnetic exchange between the
cluster entities can be varied over a wide range. Examples for
such cluster-based networks are e.~g. given by
chains\cite{MDT:ACIE02} and square
lattices\cite{MSS:ACIE00,MSS:SSS00} of the magnetic Keplerate
molecule \{$\textrm{Mo}_{72}\textrm{Fe}_{30}$\}.  These systems
show new combinations of physical properties that stem from both
molecular and bulk effects.

%===================    figure   =================================
\begin{figure}[!ht]
\begin{center}
\epsfig{file=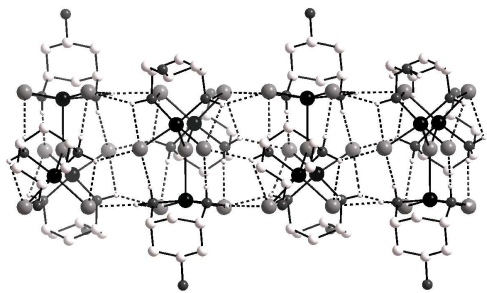,width=65mm}
\vspace*{1mm}
\caption[]{The figure shows the molecular structure of the
  cluster framework and the dotted lines represent the hydrogen
  bonds which connected the triangular units to the triangular
  anti-prismatic chain; the copper(II) ions as large black
  spheres, the nitrogen atoms as small black spheres, the
  chloride ions as large grey spheres, the carbon atoms as large
  white spheres, and the hydrogen atoms as small white spheres.}
\label{F-0}
\end{center}
\end{figure}
%===================    figure ===============================

\begin{figure}[!ht]
\begin{center}
\epsfig{file=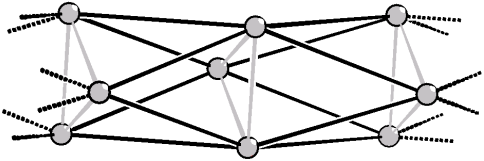,width=65mm}
\vspace*{1mm}
\caption[]{Schematic structure of the triangular copper chain:
  The copper ions are placed at the vertices, intra-triangle
  exchange pathways of the related Heisenberg Hamiltonian
  \fmref{E-3-1} are drawn by grey lines ($J_1$), inter-triangle
  couplings are given by black lines ($J_2$).}
\label{F-1}
\end{center}
\end{figure}

In this article we report the magnetic characterization of the
new cluster compound\cite{SKK:04,cu-chain}
[(CuCl$_2$tachH)$_3$Cl]Cl$_2$ (tach $=$
\emph{cis,trans}-1,3,5-triamino-cyclohexane), which contains
one-dimensional arrays of antiferromagnetically coupled
equilateral copper(II) $(s_{\text{Cu}} = 1/2)$ triangles that
are aligned to form infinite stacks of antiprisms, see
\figref{F-0}. These copper(II) triangle clusters
(CuCl$_2$tachH)$_3$Cl comprise a central $\mu_3$-chloride ion
that coordinates to the three copper centers in an unprecedented
trigonal-planar bonding mode.  In addition, this arrangement is
stabilized by hydrogen bonds between the terminal chloride
ligands and the NH groups from neighboring tach ligands
coordinating to neighboring copper centers. As the unpaired
electron of each $d^9$ Cu(II) center resides in a molecular
orbital of dominant $d_{x^2-y^2}$ character which overlap with
the ligand atoms involved in these hydrogen bonds, magnetic
superexchange between the copper centers of a triangle should be
restricted to this pathway, compare \figref{F-1}, while exchange
involving the central $\mu_3$-chloride ligand would require a
significant population of the Cu $d_{z^2}$ orbitals.  Further
hydrogen bonds between the (CuCl$_2$tachH)$_3$Cl triangles in
the crystal lattice result in a extended three-dimensional
supramolecular arrangement: additional non-coordinating chloride
anions interlink the triangles to $ab$-planes, and an extensive
network of hydrogen bonds between the terminal chloride ligands
and the NH groups connect these planes together along the
crystallographic $c$-axis.

%===================    figure   =================================
%===================    figure ===============================

Taking into account that the coupling between the
antiferromagnetically coupled triangles is mediated by
hydrogen-bonded Cu-Cl$\cdots$H-N-Cu super-exchange pathways, one
could conjecture that the magnetic properties of the system may
be well described by weakly coupled triangles. This implies that
the system at high temperatures behaves like independent
triangles while at low temperatures it behaves like a linear
chain of effective spins given by the ground state spin of each
triangle, which is $1/2$.

In a magnetization measurement such a situation would be
reflected by a pronounced plateau at $1/3$ of the saturation
magnetization. Our high-field magnetization measurement show,
however, no plateau. The observed magnetization and
susceptibility curves are successfully interpreted by a
theoretical model, in which the inter- and intra-triangle
exchange parameters are of similar magnitude. Actually, the
hydrogen-bonded super-exchange appears to be even stronger than
the intra-triangle exchange pathways (mediated by the bridging
chloro ligand and intra-molecular hydrogen bonds) although the
Cu-Cu distance between triangles is 6.82~\AA, whereas the
intra-triangle Cu-Cu distance is 4.46~\AA.

From a solid state physics point of view the new chain system
belongs to the class of Heisenberg three-leg ladder systems with
frustrated rung boundary condition, see e.~g.
Refs.~\onlinecite{KaT:JPSJ97,HMT:EPJB00,GRM:PRL03}.  Such
systems show a rich phase diagram and may have analytically
known ground states.\cite{KaT:JPSJ97} However, the three-leg
copper ladder discussed in this article is not of the simple
structure -- investigated in
Refs.~\onlinecite{KaT:JPSJ97,HMT:EPJB00,GRM:PRL03} -- where one
either assumes that the coupling is simpler\cite{KaT:JPSJ97} or
that the rung spin is a good quantum number.\cite{HMT:EPJB00}
Nevertheless, the present structure poses a new interesting
frustrated three-leg ladder with two competing interactions.

The article is organized as follows. In sec.~\ref{sec-2} we report
our experimental results followed by a discussion of the
theoretical model in sec.~\ref{sec-3}. The last section outlines
the behavior of our three-leg ladder for fictitious intra- and
inter-triangle couplings.

%%%%%%%%%%%%%%%%%%%%%%%%%%%%%%%%%%%%%%%%%%%%%%%%%%%%%%%%%%%%%%%%%%%%%%%%
\section{Experimental results}
\label{sec-2}

%===================    figure   =================================
\begin{figure}[!ht]
\begin{center}
\epsfig{file=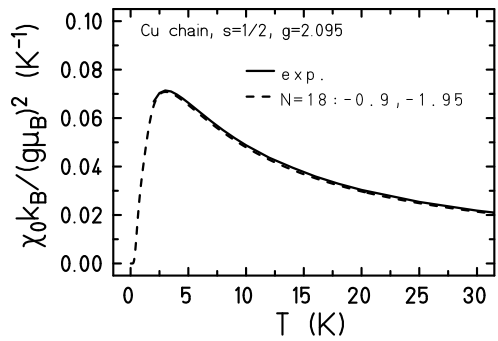,width=65mm}
\vspace*{1mm}
\caption[]{Magnetic low-field susceptibility $\chi_0$ per copper
  triangle: The solid curve shows the experimental result
  obtained at $B=0.5$~Tesla down to temperatures of 2~K. The
  dashed curve displays the zero-field susceptibility of an
  isotropic Heisenberg model, compare Eq.~\fmref{E-3-1}, with
  exchange parameters $J_1=-0.9$~K and $J_2=-1.95$~K as given in
  the figure. }
\label{F-2}
\end{center}
\end{figure}
%===================    figure ===============================

\begin{figure}[!ht]
\begin{center}
\epsfig{file=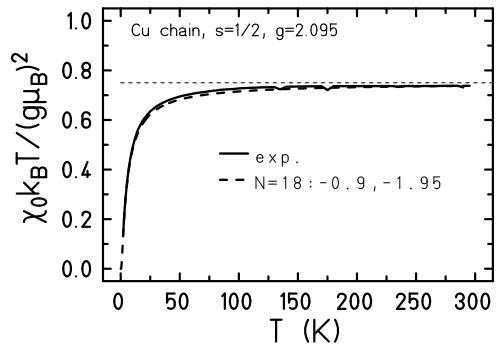,width=65mm}
\vspace*{1mm}
\caption[]{Magnetic low-field susceptibility $T \chi_0$ per copper
triangle: Again the solid curve shows the experimental result,
the dashed curve the theoretical one. The thin horizontal line
is the high temperature limit.}
\label{F-3}
\end{center}
\end{figure}

The susceptibility $\chi$ was measured for
[(CuCl$_2$tachH)$_3$Cl]Cl$_2$ using a SQUID magnetometer (Quantum
Design MPMS-5) at $B=0.5$~Tesla in a temperature range of $2 -
290$~K.  The resulting dependence is shown as a solid curve in
\figref{F-2}. Figure \ref{F-3} displays the related dependence
of $T\chi$ on temperature. Here the thin dashed line marks the
high temperature limit for an equivalent system of three
uncorrelated $s=1/2$ centers.

%===================    figure   =================================
%===================    figure ===============================

To establish the $g$-values of [(CuCl$_2$tachH)$_3$Cl]Cl$_2$ ESR
measurements have been performed at 190~GHz with a powder sample
by using pulsed magnetic fields and Gunn
oscillators.\cite{HMO:JPSJ03} A typical powder pattern has been
found. We fitted the ESR spectrum determined at 30~K, where one
can neglect the effect of short range order, compare
\figref{F-3}. The spectrum can be very well reproduced by
assuming a uni-axial anisotropy of the $g$-value with
$g_\parallel=2.06\pm 0.02$ and $g_\perp=2.12\pm 0.02$. These
$g$-values can be interpreted by considering the local
environment of Cu$^{2+}$ ions as follows.  Cu$^{2+}$ is in
pyramidical coordination and the apical bonds are in the plane
normal to the chain ($ab$-plane). The lengths of the two bonds
in the basal plane of each pyramid are nearly the same and thus
the $g$-value anisotropy in the basal plane is small. In this
case, we can define the local $g$-value along the apical bond as
$g_a$ and in the basal plane as $g_b$. Note that directions of
the apical bonds alter by $120^\circ$ between nearest neighbor
Cu$^{2+}$ ions in a triangle. Therefore, the effective $g$-value
of coupled copper ions in the $ab$-plane is the average of $g_a$
and $g_b$ with some weak three-fold anisotropy.  The $g$-value
anisotropy in the $ab$-plane for the three-fold symmetry of a
triangle cannot be seen in the present experiment due to the use
of a powder sample. If a single crystal was used, the three-fold
pattern may be observed at very high frequency ($>600$~GHz)
where the anisotropy of the Zeeman energy is much larger than
the exchange coupling. Further, the effective $g$-value along
the chain corresponds to $g_b$, because it is common for all
triangles.  Accordingly, we can expect a uni-axial $g$-value
anisotropy, where $g_b$ is close to $g_\parallel$ and $g_\perp$
is the average of $g_a$ and $g_b$.  For the theoretical studies
of section~\ref{sec-3} we are using an average $g$-value of
2.095.

%===================    figure   =================================
\begin{figure}[!ht]
\begin{center}
\epsfig{file=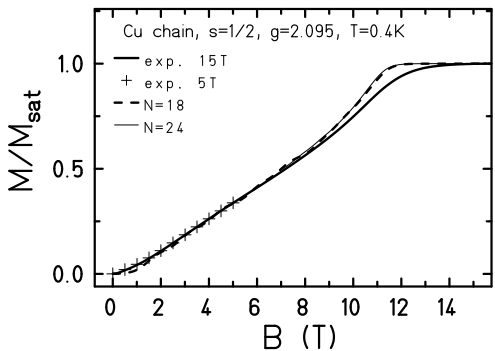,width=65mm}
\vspace*{1mm}
\caption[]{Magnetization per copper triangle: Experimental data
  taken at a cryostat temperature of 0.4~K are given by thick
  solid lines, the two data sets are practically identical. The
  theoretical estimates using the same parameters as in
  \figref{F-2} are given for 18 and 24 spins, i.~e. 6 or 8
  triangles. The saturation field is $B_{\text{sat}}=11$~Tesla.}
\label{F-4}
\end{center}
\end{figure}
%===================    figure ===============================

The high-field magnetization of [(CuCl$_2$tachH)$_3$Cl]Cl$_2$ has
been measured in pulsed magnetic fields at the Okayama high
magnetic field laboratory by using a standard inductive method
(maximum $B=40$~Tesla, $dB/dt=10000\dots 150000$~Tesla/s). The
sample is immersed in liquid $^3$He to keep a good contact with
the thermal bath.  No clear hysteresis is found between up and
down sweeps, see \figref{F-4}. In addition static-field
measurements were performed up to 5~T which perfectly agree with
the pulsed-field results. Thus we conclude that the measured
magnetization curve shows an isothermal magnetization.
Nevertheless, the experimental curve deviates from the
theoretical one, see Sec.~\xref{sec-3} below, especially in the
vicinity of the saturation field $B_{\text{sat}}=11$~Tesla.  It
might very well be that the $g$-value anisotropy or a staggered
field, which is possible in the present crystal structure, are
the origin of the stronger rounding of the experimental data.
The effect of a staggered field is strongly enhanced in higher
fields since it grows with the external field strength, see
e.~g.  Ref.~\onlinecite{SaS:JPSJ94} where a linear dependence
between staggered and external field is assumed which leads to
rounding effects (due to mixing of states) that are quadratic in
the staggered field.

The small difference between experimental and theoretical curve
at low fields is due to the fact that a finite chain has a
larger singlet-triplet gap compared to the infinite chain, see
\figref{F-7}.

%%%%%%%%%%%%%%%%%%%%%%%%%%%%%%%%%%%%%%%%%%%%%%%%%%%%%%%%%%%%%%%%%%%%%%%%
\section{Theoretical Model}
\label{sec-3}

Our EPR measurements suggest only weak anisotropy, therefore we
model the system by an isotropic Heisenberg Hamiltonian
%--------------------------------------------------------
\begin{eqnarray}
\label{E-3-1}
\op{H}
&=&
-
\sum_{u, v}\;
J_{uv}\,
\op{\vec{s}}(u) \cdot \op{\vec{s}}(v)
+
g \mu_B \vec{B} \cdot \op{\vec{S}}
\ ,
\end{eqnarray}
%--------------------------------------------------------
where $\op{\vec{s}}(u)$ are the individual spin operators at
sites $u$, $\op{\vec{S}}$ is the total spin, $\op{S}_z$ its
$z$-component. The homogeneous magnetic field defines the
$z$-direction. The spin quantum number of the copper ions is
$s=1/2$. $J_{uv}$ are the matrix elements of the symmetric
coupling matrix. A negative value of $J_{uv}$ corresponds to
antiferromagnetic coupling.

Using the fact that the Hamiltonian, the total spin, its
$z$-component as well as the point-group symmetry operations
commute with each other it is possible to diagonalize the
Hamiltonian of finite chains numerically exactly.  Thus all the
thermodynamic quantities can be evaluated without approximation.
We determined the complete spectra of two, four, and six
triangles with cyclic boundary condition in order to estimate
the unknown intra-triangle exchange parameter $J_1$ as well as
its inter-triangle counterpart $J_2$.  For eight triangles the
spectrum could still be determined for Hilbert subspaces with
large enough magnetic quantum number $M$. Since the resulting
magnetization does not differ from those of a six-triangle
chain, see \figref{F-4}, we can safely assume that a chain of
six triangles already reflects the properties of the infinite
chain to a very large extent.

The Figures \ref{F-2}, \ref{F-3}, and \ref{F-4} show the
theoretical curves for the zero-field susceptibility as well as
for the magnetization, which we obtain for $J_1=-0.9$~K and
$J_2=-1.95$~K. The magnetization curve turns out to be crucial
for an understanding of the exchange parameters. If the
intra-triangle coupling was much stronger than the
inter-triangle coupling, then the magnetization curve would show
a pronounced plateau at $1/3$ of the saturation magnetization.
This can be understood by looking at an equilateral triangle of
spins $s=1/2$. In such a triangle there are only two energy
levels, one for total spin $S=1/2$, the other for $S=3/2$, both
levels are fourfold degenerate. At $T=0$ the magnetization is
$1/2$ for low fields until one reaches the field where the
lowest Zeeman level of $S=3/2$ crosses the former ground state
level. Thus, in such a triangle the $T=0$ magnetization curve
has a long plateau and one jump to saturation, compare also
\figref{F-6}. For weakly coupled triangles this plateau starts
to shrink from both sides until it vanishes for couplings of the
same order. This situation holds for the investigated compound,
compare \figref{F-4}, therefore, this chain is an example of a
frustrated spin system with competing interactions.

%===================    figure   =================================
\begin{figure}[!ht]
\begin{center}
\epsfig{file=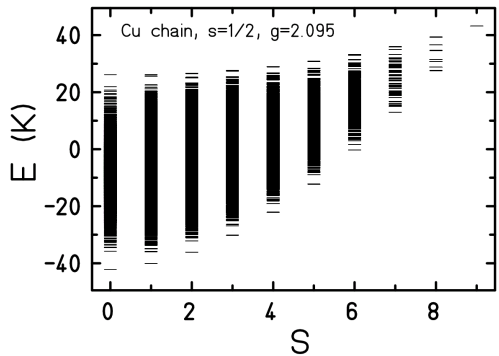,width=65mm}
\vspace*{1mm}
\caption[]{Energy eigenvalues for a finite chain of 6 triangles
  using the exchange parameters given in \figref{F-2}.}
\label{F-5}
\end{center}
\end{figure}
%===================    figure ===============================

Figure \ref{F-5} shows the energy eigenvalues as a function of
total spin for the finite chain with six triangles. The ground
state is a singlet which is clearly separated from the first
triplet excitation. There are no singlet intruders below the
triplet state which sometimes is a sign of strong
frustration.\cite{WEB:EPJB98}

%%%%%%%%%%%%%%%%%%%%%%%%%%%%%%%%%%%%%%%%%%%%%%%%%%%%%%%%%%%%%%%%%%%%%%%%
\section{Zero-temperature behavior of the new three-leg ladder}
\label{sec-4}

The magnetism of the three-lag ladder shown in \figref{F-1}
exhibits three extremes depending on $J_1$ and $J_2$. If the
inter-triangle exchange parameter $J_2$ is equal to zero the
chain decays into independent triangles, if the intra-triangle
exchange parameter $J_1$ is zero the three-lag ladder transforms
into a square lattice wrapped around a torus, and if $J_1=J_2<0$
the three-lag ladder is a triangular lattice on a torus.

From a classical point of view the ground states and the
magnetization curves for $T=0$ are analytically known for all
combinations of antiferromagnetic couplings $J_1$ and
$J_2$.\cite{Schmidt:PC} Here we will discuss the special cases
$J_1=0, J_2<0$ and $J_1<0, J_2=0$ as well as $J_1=J_2<0$.  In
the first case $J_1=0, J_2<0$ the system is bipartite and the
classical ground state shows collinear N\'eel order.  In the
other two cases the ground state is coplanar with relative
angles of $120^\circ$ between neighboring spins because of
three-colorability of the spin system.\cite{AxL:PRB01,ScL:JPA03}
It can be shown that these two states always constitute a ground
state. For $|J_1| < 2/3 |J_2|$ the collinear N\'eel state is
ground state, for $|J_1| > 2/3 |J_2|$ the $120^\circ$
arrangement becomes ground state.\cite{Schmidt:PC} For all
antiferromagnetic couplings the ground state has no chiral
structure.  In addition the minimal energies form a parabola as
a function of total spin $S$, therefore the magnetization curve
for $T=0$ increases linearly up to saturation.

%===================    figure   =================================
\begin{figure}[!ht]
\begin{center}
\epsfig{file=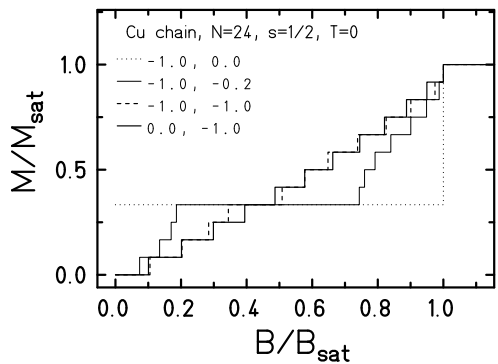,width=65mm}
\vspace*{1mm}
\caption[]{Magnetization curves for $T=0$ for various hypothetical
  couplings of a chain of eight triangles with cyclic boundary
  condition. The used combinations of $J_1$ and $J_2$ are given
  in the figure together with the respective line mode.}
\label{F-6}
\end{center}
\end{figure}
%===================    figure ===============================

In the following we discuss the behavior of a finite
quantum-spin chain of 8 triangles for various combinations of
$J_1$ and $J_2$. Figure~\ref{F-6} shows magnetization curves for
$T=0$ for various hypothetical couplings. The dotted curve shows
the magnetization of independent triangles ($J_2=0$), the dashed
curve displays the behavior if both couplings are of equal size,
and the solid curve denotes the magnetization for vanishing
intra-triangle coupling ($J_1=0$). The thin curve gives an
example for weakly coupled triangles ($J_1=1, J_2=0.2$).

The discussion of independent triangles can be kept rather
short, since this system can be analytically solved. For $s=1/2$
the ground state has $S=1/2$ and the first (and sole) excited
state has $S=3/2$. Both are fourfold degenerate. The
magnetization consists of a long plateau at $1/3$ saturation
magnetization with a single jump to saturation.

The opposite case with vanishing intra-triangle coupling is
related to the square lattice in the sense that the triangular
chain in this case is a square lattice on a torus which extends
infinitely only along one direction. In the orthogonal direction
the lattice has a periodic boundary condition with a period of
three spins. Since the system is bipartite the theorems of Lieb,
Schultz, and Mattis apply,\cite{LSM:AP61,LiM:JMP62} i.~e. the
ground state has $S=0$ and is non-degenerate.

The symmetric case with $J_1=J_2<0$ corresponds to a triangular
lattice on a torus which extends infinitely only along one
direction and again has a periodic boundary condition with a
period of three spins in the orthogonal direction, see also
Refs.~\onlinecite{Hon:JPCM99,YOS:04}.  Although it is nowadays
accepted that the triangular lattice has an ordered ground state
and gapless magnetic excitations,\cite{Cap:IJMPB01} these
properties do not automatically extend to the triangular chain
due to the different dimensionality. It is very likely that the
ground state is not ordered because the chain is quasi
one-dimensional.

The magnetization curves of finite systems for the cases $J_1=0,
J_2<0$ and $J_1=J_2<0$ look rather featureless, resembling an
almost regular staircase with no prominent plateau or jump.  The
cases interpolating between uncoupled and coupled triangles show
a plateau at $1/3$ saturation magnetization, which is longer or
shorter depending on the ratio of inter- and intra-triangle
coupling.

%===================    figure   =================================
\begin{figure}[!ht]
\begin{center}
\epsfig{file=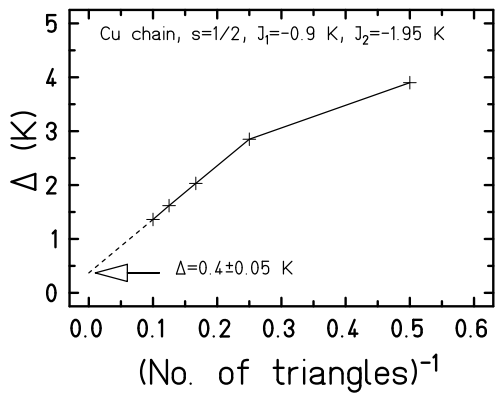,width=65mm}
\vspace*{1mm}
\caption[]{Finite size extrapolation of the singlet-triplet gap
  for [(CuCl$_2$tachH)$_3$Cl]Cl$_2$. The gap for the infinite
  chain is estimated to be $\Delta\gtrapprox 0.4\pm0.05$~K.}
\label{F-7}
\end{center}
\end{figure}
%===================    figure ===============================

After having discussed the properties of finite chains it is of
course desirable to deduce properties of the infinite chain
under investigation. For [(CuCl$_2$tachH)$_3$Cl]Cl$_2$ the
ground state and the singlet-triplet gap can be evaluated using
a Lanczos procedure for systems up to 10 triangles. All finite
size calculations result in a ground state spin $S=0$ which is
separated by a finite gap from the triplet state, compare
\figref{F-5}. Using chains with up to 10 triangles enables us to
extrapolate to the infinite system, which turns out to have a
ground state spin $S=0$ and a non-zero gap of $\Delta\gtrapprox
0.4\pm0.05$~K, compare \figref{F-7}.  The value of the gap is
very likely somewhat bigger since one can expect that the
convergence changes from linearly in $1/N$ to
exponential\cite{GeH:JPA93,GJL:PRB94} or a power
law\cite{CHP:PRB98} like $1/N^2$ once the system size is bigger
than the correlation length $\xi$.

%===================    figure   =================================
\begin{figure}[!ht]
\begin{center}
\epsfig{file=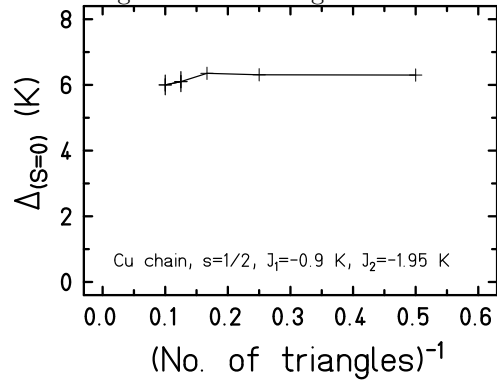,width=65mm}
\vspace*{1mm}
\caption[]{Finite size behavior of the singlet-singlet gap
  for [(CuCl$_2$tachH)$_3$Cl]Cl$_2$. The two leftmost data
  points for eight and ten triangles have an uncertainty of
  about $\pm0.2$~K due to the restricted accuracy of the Lanczos
  procedure for excited states.}
\label{F-8}
\end{center}
\end{figure}
%===================    figure ===============================
The ground state of the [(CuCl$_2$tachH)$_3$Cl]Cl$_2$ chain is
non-degenerate for all simulated system sizes. This behavior
differs from the behavior of chains of weakly coupled triangles
where a dimerization together with a twofold degeneracy of the
ground state is observed.\cite{LNM-CM04} With very high
confidence we can also state that the big singlet-singlet gap --
compare \figref{F-5} -- does not collapse to a degenerate ground
state with increasing system size, see \figref{F-8}.  The gap is
practically independent of $N$. The two values for eight and ten
triangles have an uncertainty of about $\pm0.2$~K due to the
insufficient accuracy of the Lanczos procedure for such
high-lying levels. We thus regard the slight decrease for these
values as not significant.

%%%%%%%%%%%%%%%%%%%%%%%%%%%%%%%%%%%%%%%%%%%%%%%%%%%%%%%%%%%%%%%%%%%%%%%%
\section*{Acknowledgments}

We thank Peter Hage for letting us utilize his Lanczos program,
and we thank Andreas Honecker, Johannes Richter, Heinz-J\"urgen
Schmidt as well as J\"org Schulenburg for fruitful discussions
and again Heinz-J\"urgen Schmidt for carefully reading the
manuscript.  H.~N. acknowledges the support by Grant in Aid for
Scientific Research on Priority Areas (No. 13130204) from MEXT,
Japan and by Shimazu Science Foundation. The Ames Laboratory is
operated for the United States Department of Energy by Iowa
State University under Contract No. W-7405-Eng-82.

%\bibliography{/home/schnack/tex/bibtex/js-own,/home/schnack/tex/bibtex/js-mag,/home/schnack/tex/bibtex/js-mis}

\end{document}